\pdfoutput=1
\documentclass[aps,prl,twocolumn,superscriptaddress]{revtex4-1}
\usepackage{graphicx,color} 
\usepackage{dcolumn}
\usepackage{bm}
\usepackage{amssymb,amsmath}
\usepackage{epstopdf}

\usepackage{upgreek}
\usepackage[normalem]{ulem}

\newcommand{\be}{\begin{equation}}
\newcommand{\ee}{\end{equation}}

\newcommand{\ii}{\ensuremath{\mathrm i}}
\newcommand{\dd}{\ensuremath{\mathrm d}}

\renewcommand{\pi}{\uppi}
\renewcommand{\Psi}{\varPsi}
\renewcommand{\delta}{\updelta}
\renewcommand{\k}{\mathbf{k}}
\newcommand{\q}{\mathbf{q}}
\renewcommand{\l}{\mathbf{l}}
\newcommand{\K}{\mathbf{K}}
\newcommand{\x}{\mathrm{X,CM}}
\newcommand{\qkk}{{\q,\k',\k}}
\newcommand{\ddt}{\frac\partial{\partial t}}
\newcommand{\e}{\mathrm e}
\newcommand{\h}{\mathrm h}
\newcommand{\I}{\mathrm i}

\newcommand{\X}{\mathrm{X}}
\newcommand{\ba}{\begin{eqnarray}}
\newcommand{\ea}{\end{eqnarray}}
\newcommand{\bi}{\begin{itemize}}
\newcommand{\ei}{\end{itemize}}

\begin{document}
\title{Observation of Coulomb-Assisted Dipole-Forbidden Intraexciton Transitions \\in Semiconductors
}

\normalsize
\author{W.~D.~Rice,$^{1,2}$ J.~Kono,$^{1,2,*}$ S.~Zybell,$^{3,4}$ S.~Winnerl,$^3$ J.~Bhattacharyya,$^3$ H.~Schneider,$^3$ 
M.~Helm,$^{3,4}$ B.~Ewers,$^5$ A.~Chernikov,$^5$ M.~Koch,$^5$ S.~Chatterjee,$^5$ G.~Khitrova,$^6$ H.~M.~Gibbs,$^{\dag, 6}$
L.~Schneebeli,$^5$ B.~Breddermann,$^5$ M.~Kira,$^5$ and S.~W.~Koch,$^5$\\
\small{\textit{$^1$Department of Electrical and Computer Engineering, Rice University, Houston, Texas 77005, USA}} \\
\small{\textit{$^2$Department of Physics and Astronomy, Rice University, Houston, Texas 77005, USA}} \\
\small{\textit{$^3$Helmholtz-Zentrum Dresden-Rossendorf, P.O.~Box 510119, D-01314 Dresden, Germany}} \\
\small{\textit{$^4$Technische Universit\"at Dresden, 01062 Dresden, Germany}} \\
\small{\textit{$^5$Department of Physics, Philipps University Marburg, Renthof 5, D-35032 Marburg, Germany}} \\
\small{\textit{$^6$College of Optical Science, University of Arizona, Tucson, Arizona 85721-0094, USA}} \\
$^*$corresponding author: kono@rice.edu
}

\begin{abstract}
We use terahertz pulses to induce resonant transitions between the eigenstates of optically generated exciton populations in a high-quality semiconductor quantum-well sample. Monitoring the excitonic photoluminescence, we observe transient quenching of the $1s$ exciton emission, which we attribute to the terahertz-induced $1s$-to-$2p$ excitation. Simultaneously, a pronounced enhancement of the $2s$-exciton emission is observed, despite the $1s$-to-$2s$ transition being dipole forbidden. A microscopic many-body theory explains the experimental observations as a Coulomb-scattering mixing of the 2$s$ and 2$p$ states, yielding an effective terahertz transition between the 1$s$ and 2$s$ populations.
\end{abstract}

\pacs{78.55.-m, 71.35.-y, 78.67.De}

\maketitle



Coulombically bound electron-hole pairs, i.e., excitons, often dominate the optical properties \cite{SQObook2011} of high-quality semiconductors at low temperatures. In direct-gap semiconductors, the intra-exciton energy spacing is typically in the terahertz (THz) frequency range, with the corresponding intra-excitonic transitions following the same selection rules as the dipole transitions in hydrogen atoms. Therefore, the presence of exciton populations can be unambiguously detected by monitoring the $1s$-to-$2p$ transition resonance \cite{TimuskSolStateComm1978,LabriePRL1988,HodgePRB1990,GroeneveldOptSocAmB1994,Kira2001,KaindlNature2003,KiraSolidStateComm2004,GalbraithPRB2005,Suzuki2009}. One can also use strong THz pulses to induce nonlinear $1s$-to-$2p$ transitions \cite{Nordstrom1998,KubouchiPRL2005, HuberPRL2006, Danielson2007, LeinssPRL2008}, including Rabi flopping \cite{Luo2004,Carter2005,WagnerPRL2010, Ewers2012} and excitations in $\Lambda$ systems \cite{Tomaino2012}. 

An elegant way to study the influence of THz-induced exciton transitions is to monitor time-resolved photoluminescence (PL) spectra after combined optical and THz excitations \cite{Ulbricht2011}. Since only the optically active $s$ states contribute to the PL, the THz-induced intra-excitonic $1s$-to-$2p$ population transfer can be observed as  pronounced quenching of the $1s$ PL \cite{CernePRL1996,SalibPRL1996,KonoPRL1997}. Similar PL quenching can be observed for quantum wells (QWs) when the THz energy is resonant with a dipole-allowed intersubband transition~\cite{ZybellAPL2011}. 
\begin{figure}
\includegraphics [width=0.9 \columnwidth] {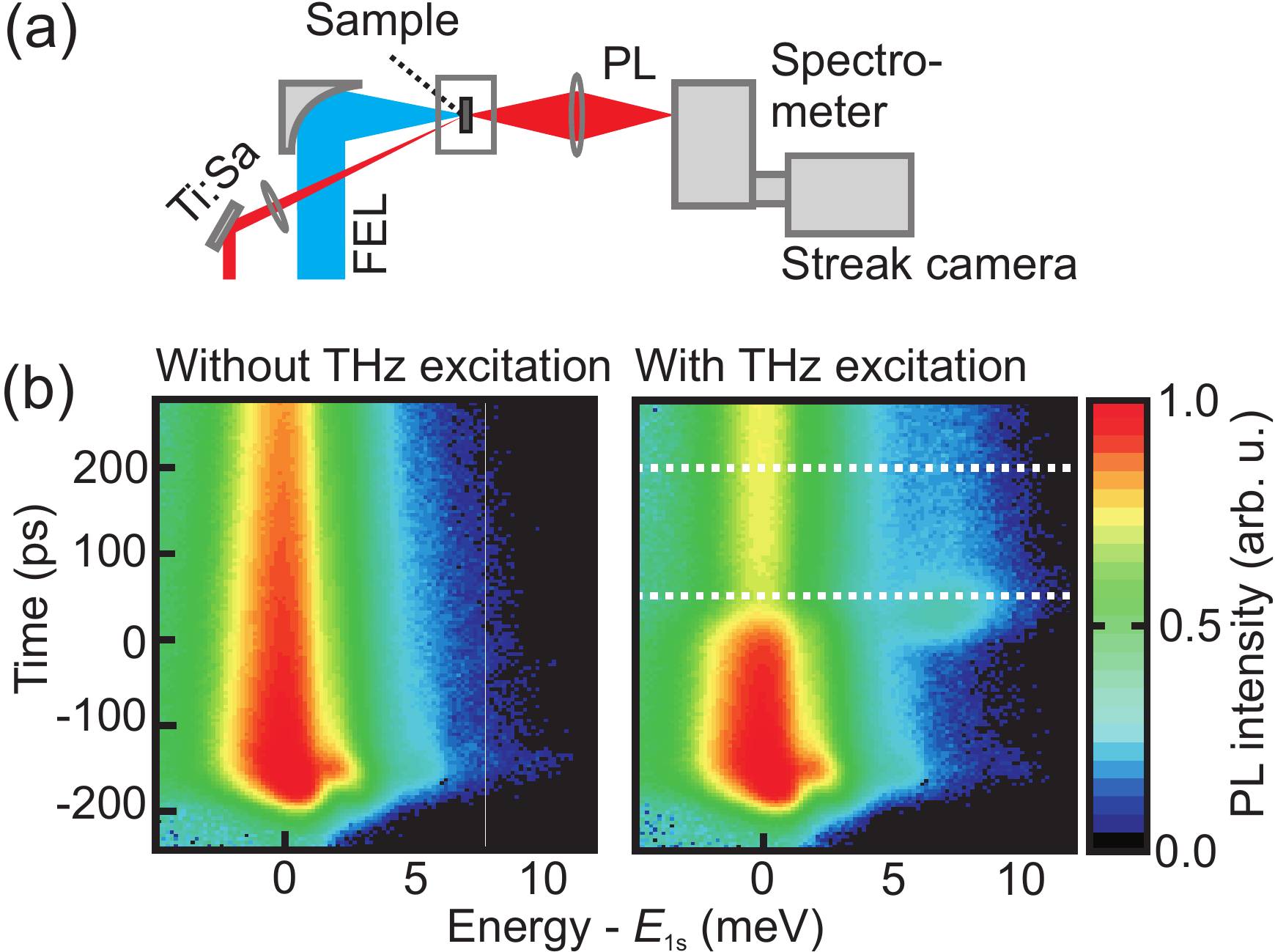}
\includegraphics [width=0.9 \columnwidth] {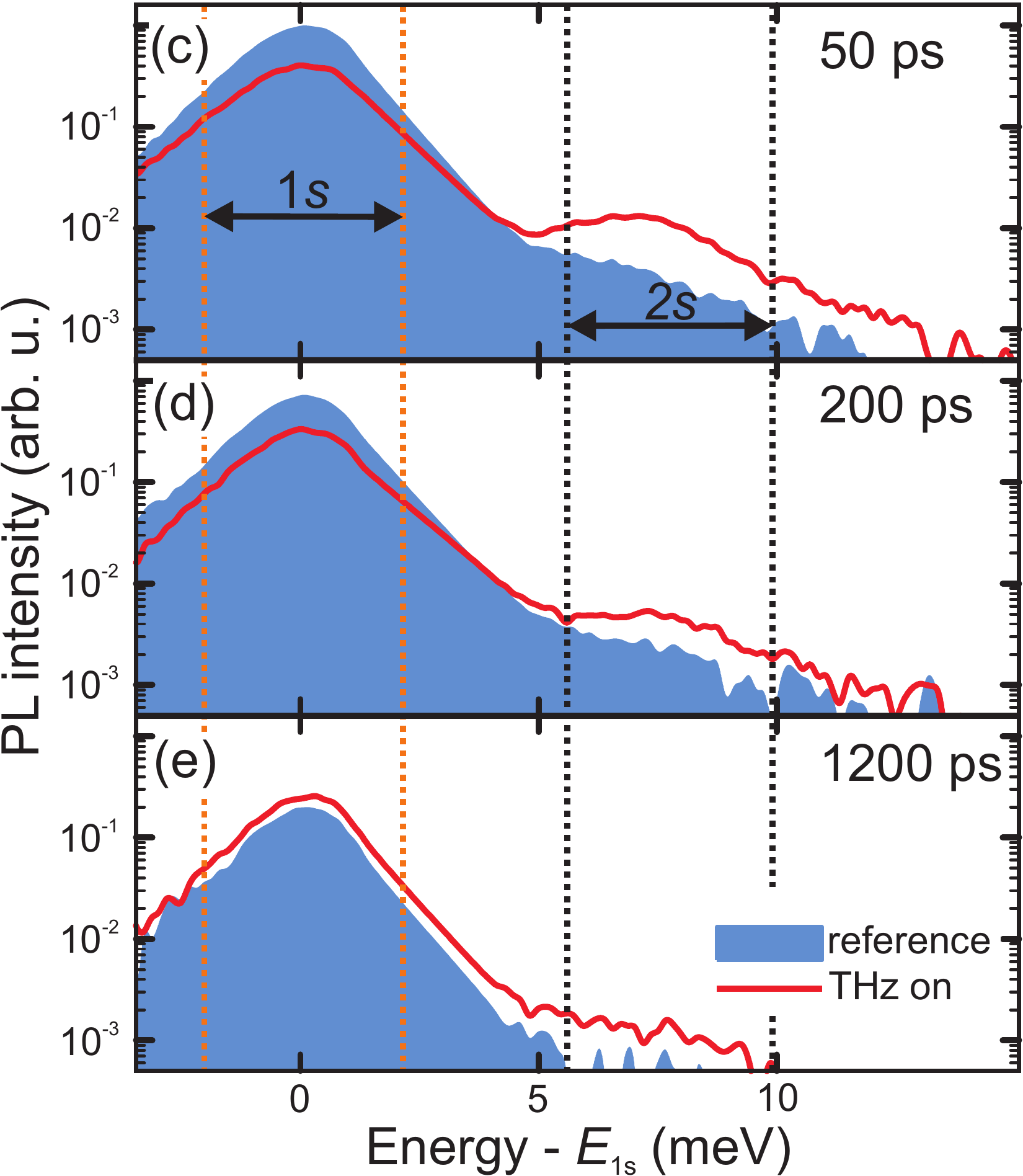}
 \caption{(color online).
 (a) Schematic experimental setup.
 (b) False-color representation of the temporally and spectrally resolved PL without (left) and with (right) THz excitation.
 (c)-(e) Measured reference PL spectra (shaded area) vs.~THz-on PL (solid red line) on a semilogarithmic scale defined at different times $t$ after the THz excitation indicated by the horizontal dashed lines in panel (b).
 }
\label{Fig:setup_dyn}
\end{figure}

In this Letter, we apply this technique to monitor how a THz pulse changes the time-resolved PL long after a resonant optical pulse has generated exciton population in the system. Besides $1s$-PL quenching, we observe an unexpected transient increase of the $2s$-PL indicating a pronounced effective $1s$-to-$2s$ transition. Using a systematic many-body theory~\cite{PQE2006}, we show that the THz-induced $1s$-to-$2p$ excitation is accompanied by an efficient 2$p$-to-2$s$ transition which can be attributed to co-operative Coulomb scattering. This Coulomb-assisted THz-induced $1s$-to-$2s$ coupling is unique to interacting many-body semiconductor configurations and cannot be observed in atomic systems.


In our experiments, we study the intra-excitonic transitions in a 20$\times$ In$_{0.06}$Ga$_{0.94}$As QW structure at a lattice temperature of 10\,K. The 8\,nm wide QWs are separated by 130\,nm GaAs barriers, rendering all nontrivial coupling effects negligible~\cite{Hubner1999}. As such, the structure acts as a single QW whose PL magnitude is additively enhanced by the number of QWs~\cite{Schafer2006}. The heavy-hole (hh) excitons are well separated from both the continuum and the light-hole states allowing us to address the resonances individually and efficiently yielding a two-band situation for the hh1~{\cite{ChatterjeePRL2004,Grunwald2008}}.
For this particular In-Ga ratio, the energy difference between the $1s$ and $2s$ hh1 excitons is estimated as ${6.7}$\,meV and the $1s$-to-$2p$ transition energy is ${6.9}$\,meV. 

We monitor the time-resolved PL after an excitation sequence where a THz pulse of a free-electron laser (FEL) follows an optical pulse of a Ti:Sapphire laser, as shown schematically in Fig.~\ref{Fig:setup_dyn}(a).
The Ti:Sapphire laser emits pulses with a duration of 4\,ps (FWHM) at a repetition rate of 78\,MHz with a photon energy of 1.471\,eV which is in resonance with the 1$s$ hh state. The repetition rate is reduced to 13\,MHz by an extra-cavity pulse picker equipped with an acousto-optic modulator to match the repetition rate of the FEL.
The FEL emits 30\,ps pulses with a wavelength of 191\,$\mu$m which is in resonance with the $1s$-to-$2p$ hh exciton transition energy.
The peak field strength of the FEL beam at the position of the sample is estimated to be 5\,kV\,cm$^{-1}$.

The two sources are synchronized electronically and their time delay is controlled by using a mechanical delay of the synchronizing pulses; for details, see~\cite{BhattacharyyaRSI2011}.
The two beams are collinearly polarized and focussed directly onto the QW.
The spot sizes are chosen to be 300 and 50\,$\mu$m for the FEL and the Ti:Sapphire laser, respectively.
The PL is collected in a forward-scattering geometry through the transparent GaAs substrate.
Only the center spot of about 20\,$\mu$m in diameter is imaged within a small solid angle, carefully avoiding the transmitted laser beams as well as density-averaging effects.
The PL is spectrally and temporally dispersed using a spectrometer attached to a streak camera with energy and time resolutions of 0.15\,meV and 15\,ps, respectively; the time is oversampled, collecting a spectrum every 4.4\,ps.

The relatively weak optical pulse, that is $\approx~10^{11}$~photons$/(\mathrm{cm}^2)$ per pulse, couples to the $1s$ polarization which is converted into incoherent $1s$ excitons.
This polarization-to-population conversion process occurs efficiently with a characteristic time scale of $<$10\,ps, as shown in Ref.~\cite{KiraPRL2004}.
Exemplary PL data without THz excitation are given in the left panel of Fig.~\ref{Fig:setup_dyn}(b) where the spectrally and temporally emitted PL intensities are plotted in false colors.
The time $t$ is defined with respect to the center of the THz pulse.
Following the initial excitation, the PL decays exponentially on a time scale of 600\,ps.
The corresponding data for dual excitation are shown in the right panel.
The THz pulse arrives 225\,ps after the optical pulse, thus effectively exciting the incoherent $1s$ population into the $2p$ state and partially into the ionization continuum \cite{Ewers2012}.
This population transfer quenches the 1$s$ population and hence the related 1$s$ resonance in PL as predicted in Refs.~\cite{CernePRL1996,SalibPRL1996}.
Additionally, a clear spike in the $2s$ emission is visible when the FEL is incident, and the 1s PL recovers on a much longer time scale.

To better quantify the THz-induced changes, we plot emission spectra for three representative $t$ with (solid line) and without (shaded area)  THz excitation in Figs.~\ref{Fig:setup_dyn}(c)-\ref{Fig:setup_dyn}(e). For a short time delay of $t\,=\,50$\,ps just after the THz pulse [panel (c)], the $1s$ PL intensity decreases by 46\,\%, verifying the usual THz-field-induced quenching scenario.
However, in addition, the THz excitation induces a pronounced $2s$ resonance; this observation is unexpected because the direct 1$s$-to-2$s$ transition is dipole forbidden.
The 1$s$ PL then recovers its intensity gradually while the 2$s$ peak decays, as seen in (d) and (e).
For $t=1200$\,ps, we observe another interesting feature: the $1s$ PL becomes larger with than without THz, indicating that more luminescing $1s$ excitons are present in the system long after the THz excitation.
This is explained by THz-induced shelving of the overall exciton populations into optically dark states that cannot recombine radiatively. As the excitations relax back to the 1$s$ state, we eventually observe excess 1$s$ PL at later times because excitons experience a reduced overall radiative decay during the THz excitation-relaxation cycle.
Note that nonradiative recombination in these samples is negligible for the chosen excitation conditions as the overall time-integrated emission intensities with and without THz excitation match within the experimental error.

\begin{figure}
\includegraphics [width = 0.88 \columnwidth] {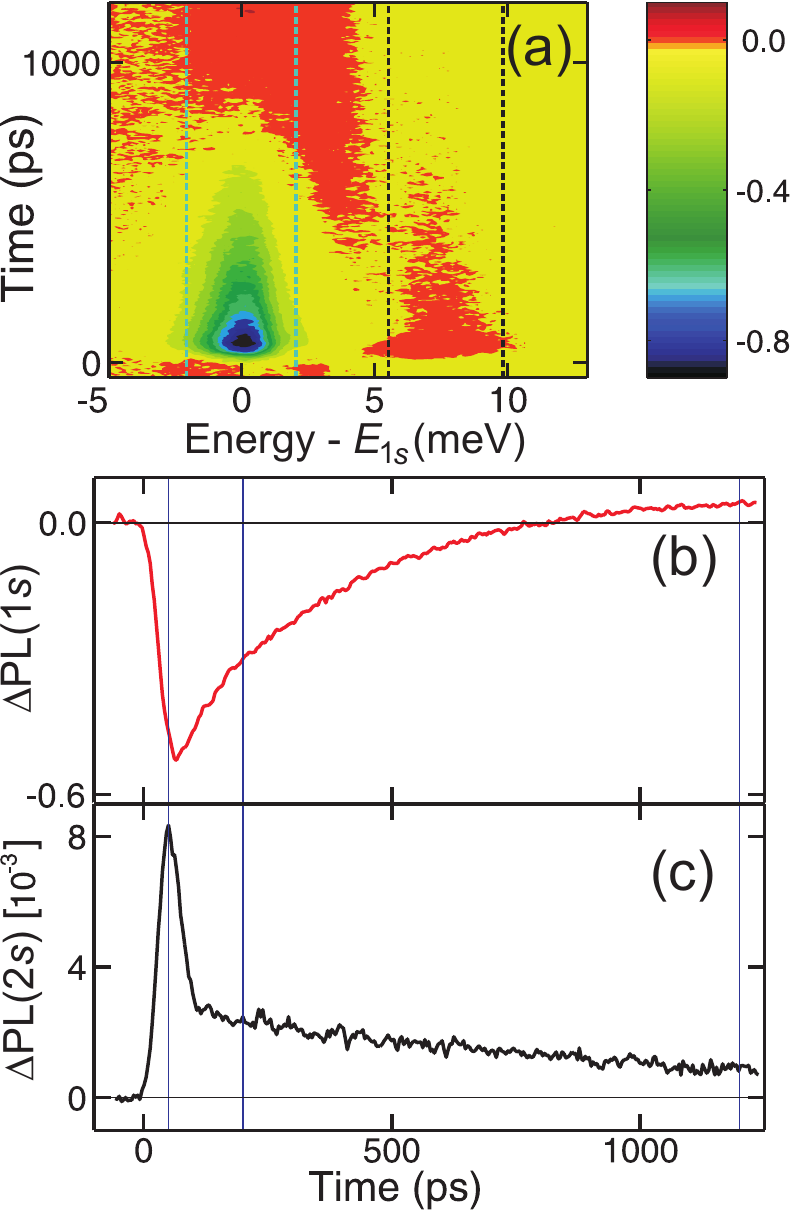}
 \caption{(color online).
 THz-induced effect on PL.
 (a) Measured differential PL spectrum.
 The vertical dashed lines mark the regions of energies used for (b) $\Delta \mathrm{PL}_{1s}$ and (c) $\Delta \mathrm{PL}_{2s}$.
 The solid vertical lines indicate the time slices shown in Figs.~\ref{Fig:setup_dyn}(c)-\ref{Fig:setup_dyn}(e).
}
\label{Fig:TRPLdiffE}
\end{figure}

We follow the time evolution of these THz-induced phenomena by determining the differential photoluminescence
$\Delta \mathrm{PL} (t) = {\rm PL}_{\rm{THz}} - {\rm PL}_{\rm ref}$ between the cases with (${\rm PL}_{\rm{THz}}$) and without THz  excitation (${\rm PL}_{\rm ref}$).
The measured $\Delta \mathrm{PL}$ spectra are shown in Fig.~\ref{Fig:TRPLdiffE}(a) as a contour plot.
Again, the 1$s$ quench (early times), $1s$ shelving (later times), and the 2$s$-excess PL are clearly visible.
To monitor $\Delta \mathrm{PL}$ dynamics in more detail, Figs.~\ref{Fig:TRPLdiffE}(b) and \ref{Fig:TRPLdiffE}(c) present  $\Delta \mathrm{PL}_{1s}$ and $\Delta \mathrm{PL}_{2s}$, corresponding to energy-integrated spectra around the indicated 1$s$ and 2$s$ energies, respectively, as a function of time.
The strong negative dip in $\Delta \mathrm{PL}_{1s}$ results from the $1s$ quench while long-time $\Delta \mathrm{PL}_{1s}$ overshoots to a positive range, demonstrating the $1s$ shelving.
The THz pulse induces a strictly positive $\Delta \mathrm{PL}_{2s}$: it increases rapidly after the THz excitation, peaking at $t =$ 50 ps. Moreover, we observe no appreciable delay between the 1$s$ quench and increased 2$s$ PL. Hence, THz can induce a ``direct" 1$s$-to-2$s$ transition via the diffusive Coulomb scattering that effectively mixes the 2$s$ and 2$p$ state (see SM).
After that, $\Delta \mathrm{PL}_{2s}$ displays a double-decay behavior: a fast decay is followed by a slow exponential decay with roughly a 900\,ps relaxation constant. This shows that the THz-induced 2$s$ PL is a fast transient compared with the relaxation time scale. 


To explain the unexpected 2$s$ PL increase, we theoretically study how the Coulomb interaction modifies the THz-induced exciton transitions and the related PL. For this purpose, we concentrate on the excitation dynamics within one of the QWs. The
microscopic properties of the excitons are defined by the two-particle correlations $c_{\X}^{\qkk}\equiv \Delta \langle e_{\k}^{\dagger} h_{\k -\q}^\dagger h_{\k'} e_{\k'+ \q}   \rangle$ between the electron $e^\dagger e$ and hole $h^\dagger h$ operators~\cite{PQE2006}.
The corresponding electron (hole)  distribution is
$f_{\k}^{\e} = \langle e_{\k}^\dagger e_{\k}\rangle$ ($f_{\k}^{\h} = \langle h_{\k}^\dagger h_{\k} \rangle$).
In the so-called main-sum approximation \cite{PQE2006,SQObook2011}, the $c_{\X}$ dynamics is given by
\begin{align}
  &\I \hbar \ddt c_{\X}^{\qkk} = E_{\k,\k',\q}^{\mathrm{eh}}\, c_{\X}^{\qkk}
  - \bm{A}_{\mathrm{THz}} (t) \cdot \bm{j}_{\k'+ \q -\k} \, c_{\X}^{\qkk}
\notag
\\
& \qquad  + (1- f_{\k}^{\e} -f_{\k-\q}^{\h}) \sum_{\l} V_{\l-\k} \, c_{\X}^{\q, \k', \l}
\notag
\\
& \qquad - (1- f_{\k'+\q}^{\e} -f_{\k'}^{\h}) \sum_{\l} V_{\l-\k'} \, c_{\X}^{\q, \l, \k}
 +  T^{\qkk}  ,
\label{eq:corrdyn}
\end{align}
where $E_{\k,\k',\q}^{\mathrm{eh}}$ contains the renormalized energy of an electron-hole pair, $V_{\k}$ is the Coulomb-matrix element, $\bm{A}_{\mathrm{THz}} (t)$ is the vector potential of the THz pulse, and $\bm{j}_{\k} = -|e| \hbar \k/\mu$ is the current matrix element with the reduced electron--hole mass $\mu$. The three-particle correlations are symbolically denoted by $T^{\qkk}$, see the Supplemental Material (SM) for more details.

As shown in Ref.~\cite{PQE2006}, $T$ is dominantly built up via the Boltzmann-type Coulomb scattering where exciton correlations exchange momentum with the plasma and the other two-particle correlations. Consequently, $T$ becomes a complicated functional of exciton populations due to the quantum kinetics involved. However, one does not need to determine the full quantum kinetics explicitly to explain the consequence of the $T$-related Coulomb scattering on THz transitions.
Instead, we only need to consider that the incoming excitons scatter into new momentum states with a constraint that the number of incoming and outgoing $c_{\X}$ correlations remains constant. In other words, $T$ generates {\it diffusive} Coulomb scattering among excitons, as shown in the SM. Therefore, it is clear that an exciton correlation --- created in the state $c_{\X}^{\qkk}$ by THz transitions --- can scatter to a new state such as $c_{\X}^{\q,\k'+\K,\k}$ or $c_{\X}^{\q,\k',\k+\K}$ where $\K$ is a typical scattering wave vector.

One can explain the consequences of the diffusive Coulomb scattering on THz transitions using an ansatz
\begin{equation}
T_{\mathrm{diff}}^{\qkk} = -\ii \hbar \gamma \left[ c_{\X}^{\qkk} - \frac{1}{2\pi} \int_0^{2\pi} \dd \theta_{\K} \, c_{\mathrm{X}}^{\bm{\mathrm{q}},\bm{\mathrm{k'+K}},\bm{\mathrm{k+K}}} \right]
\label{eq: diffusive model}
\end{equation}
where $\theta_{\K}$ is the direction of the scattering $\K$ that is assumed to have a constant magnitude and $\gamma$ defines the overall scattering strength. The introduced $T_{\mathrm{diff}}$ is a generalization of the diffusive model~\cite{KiraPRL2004} that explains the principal effects of excitation-induced dephasing~\cite{PQE2006} beyond the constant-dephasing approximation.

To determine the effect of diffusive Coulomb scattering on the THz-generated 2$p$ populations, we use the exciton transformation
$\Delta N_{\q}^{\lambda , \nu} = \sum_{\k,\k'} \phi_{\lambda}(\k) \, \phi_{\nu}^{\star}(\k')  \, c_{\x}^{\qkk}$
%
%
where $\phi_{\lambda}(\k)$ is the exciton wave function and $c_{\x}^{\qkk}$ is the center-of-mass representation of $c_{\X}$ \cite{PQE2006}.
The specific exciton populations are given by the diagonal elements $\Delta N_{\q}^{\lambda , \lambda}$ while the off-diagonal elements $\Delta N_{\q}^{\lambda,\nu\neq \lambda}$ define the exciton transitions. The diffusive scattering from 2$p$ to 2$s$ can be deduced by projecting Eqs.~(\ref{eq:corrdyn})--\eqref{eq: diffusive model} with the exciton transformation and following the 2$p$ contributions, yielding
$\frac{\partial}{\partial t} \Delta N_{\bm{\mathrm{q}}}^{2s,2s} \left. \right|_{2p} =
\Delta N_{\bm{\mathrm{q}}}^{2p,2p}/\tau_{\mathrm{conv}}$ (see SM) where we have defined a scattering time
\begin{equation}
\frac{2\pi}{\tau_{\mathrm{conv}}} \equiv
\gamma \int_0^{2\pi} \dd \theta_{\K} \,
  \left|
  {\textstyle
      \sum_{\bm{\mathrm{k}}} \phi_{2p}(\bm{\mathrm{k}}) \, \phi_{2s}^{\star} (\bm{\mathrm{k+K}})
  }
  \right|^2 .
\end{equation}
We see that the 2$p$-to-2$s$ coupling is present as long as ${\bf K}$ is not zero because the 2$s$ and 2$p$ states are orthogonal only for ${\bf K} = 0$.

Besides the 2$p$-to-2$s$ coupling, the Coulomb interaction also relaxes 2$s$ populations toward the quasi-equilibrium on a time scale $\tau_{\rm rel}$. The combined effect of $\tau_{\rm conv}$ and $\tau_{\rm rel}$ creates new Coulomb-mediated eigenstates where $2s$ and $2p$ state become mixed, see SM. In particular, the Coulomb-induced state mixing induces an effective THz transition between the original $1s$ and $2s$ states. Note that the dipole-allowed 1$s$-to-2$p$ and 2$p$-to-2$s$ transitions cannot generate efficient 1$s$-to-2$s$ population conversion without Coulomb scattering because the THz pulse is off-resonant with the 2$p$-to-2$s$ transition. 

Due to the scattering nature of $2s$-$2p$ mixing, the created $2s$ population decays with rate $\tau_{2s}^{-1}=\tau_{\rm rel}^{-1}-\tau_{\rm conv}^{-1}$ which defines also the fast decay of the excess $2s$ PL, as shown in the SM. For the late times, the $2s$ population reaches a quasi-equilibrium, yielding a slower decay of the $2s$ PL on the time scale of the remaining phonon relaxation $\tau_{\rm phon}$.

We numerically solve THz dynamics~(\ref{eq:corrdyn}) including all the relevant exciton states for the optically bright and dark excitons, and the diffusive Coulomb scattering (\ref{eq: diffusive model}). The center-of-mass momentum of dark and bright excitons is fully taken into account. We also include the radiative decay of bright excitons as well as the relaxation of excitons toward the thermodynamic equilibrium on a $\tau_{\rm phon}=900\,$ps time scale, agreeing well with independent microscopic computations \cite{Kira2001,PQE2006}. The diffusive Coulomb scattering is chosen to give $\tau_{2s}=120\,$ps ($\tau_\mathrm{conv}=56.0\,$ps, $\tau_\mathrm{rel}=38.2\,$ps) that is substantially faster than $\tau_{\rm phon}$. The quasi-stationary PL spectra are computed via the PL-Elliott formula \cite{SQObook2011}, as discussed in the SM.

\begin{figure}[tp]
\includegraphics [width = 0.88 \columnwidth] {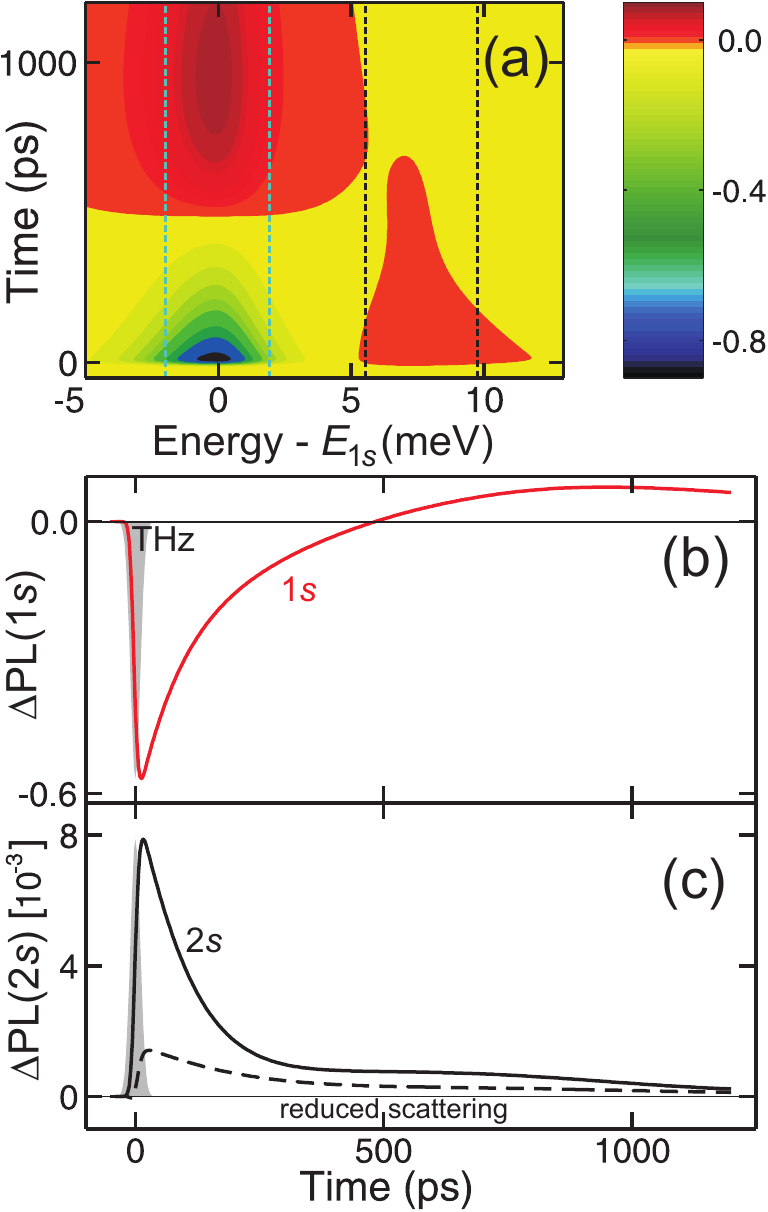}
\caption{(color online). THz-induced effect on PL. (a) Computed differential PL. Integrated differential $\Delta {\rm PL}_\lambda(t)$ for (b) $\lambda=1s$ and (c) $\lambda =2s$. The conditions are same as in Fig.~\ref{Fig:TRPLdiffE}.}
\label{Fig:TRPLdiffT}
\end{figure}

Figure~\ref{Fig:TRPLdiffT} shows the PL changes induced by THz excitation when the diffusive Coulomb scattering is included under the same excitation conditions as in Fig.~\ref{Fig:TRPLdiffE}. The computations not only explain the qualitative behavior of 1$s$ quench, exciton shelving, and excess 2$s$ PL, but they quantitatively determine the 1$s$ quench and 2$s$ excess levels. They also explain the double-decay of the $2s$ PL as switching from the fast Coulomb-equilibration $\tau_{2s}$ to the slow phonon-relaxation $\tau_{\rm phon}$. We also have performed a computation where we reduce $1/\tau_{\mathrm{conv}}$ by a factor of 20 in the $c_{\X}$ dynamics. The dashed line in Fig.~\ref{Fig:TRPLdiffT}(c) compares this
$\Delta \mathrm{PL}_{2s}$ with the full result (solid line). We see that the full and $1/\tau_{\mathrm{conv}}$-reduced computations decay similarly at the late times; note that some excess 2$s$ PL remains due to thermal relaxation from the ionized excitons toward the 1$s$ and 2$s$ states. However, only the full computation produces a strong $\Delta \mathrm{PL}_{2s}$  transient that decays fast. Hence, the $\Delta \mathrm{PL}_{2s}$ transient does not originate from relaxation but follows from the Coulomb-induced population transfer and the subsequent equilibration. The $\Delta \mathrm{PL}_{2s}$ peak also emerges on a time scale similar to the THz excitation, determined by the THz-pulse duration (here 30\,ps). Thus, the Coulomb interaction co-operates with the THz excitation to open a new 1$s$-to-2$s$ transition that is much faster than the relaxation processes.

In conclusion, our experiment-theory analysis shows the existence of Coulomb-assisted THz transitions converting $1s$ into $2s$ excitons, i.e., a process that extends the dipole-selection rules as a direct consequence of many-body interactions.
This effect survives even when an appreciable amount of disorder is present, as shown in the SM.
The related $2p$-to-$2s$ transfer is significantly faster than other relaxation processes, making the Coulomb-induced scattering an active partner in the THz transitions.
This work not only highlights a pronounced difference between excitons and atoms, but also opens up a new mechanism to manipulate excitons through combining many-body and THz effects.

\begin{acknowledgments}
This work was supported, in part, by the W.~M.~Keck Program in Quantum Materials at Rice University (W.D.R.).
The Marburg group thanks for financial support by the Deutsche Forschungsgemeinschaft.
\end{acknowledgments}


%


%
\end{document}